# Investigating Nonlinear Quenching Effects on Polar Field Buildup in the Sun Using Physics-informed Neural Networks

Jithu J. Athalathil[1], Mohammed H. Talafha[2], and Bhargav Vaidya[1]
[1] Department of Astronomy, Astrophysics and Space Engineering, Indian Institute of Technology Indore, Indore 453552, India
[2] Research Institute of Science and Engineering, University of Sharjah, 27272, Sharjah, UAE
Received 2025 November 7; revised 2026 January 10; accepted 2026 February 14; published 2026 March 13

## Abstract

The solar dynamo relies on the regeneration of the poloidal magnetic field through processes strongly modulated by nonlinear feedback such as tilt quenching (TQ) and latitude quenching (LQ). These mechanisms play a decisive role in regulating the buildup of the Sun's polar field, and in turn, the amplitude of future solar cycles. In this work, we employ physics-informed neural networks (PINNs) to solve the surface flux transport (SFT) equation, embedding physical constraints directly into the neural network framework. By systematically varying transport parameters, we isolate the relative contributions of TQ and LQ to polar dipole buildup. We use the residual dipole moment as a diagnostic for cycle-to-cycle amplification and show that TQ suppression strengthens with increasing diffusivity, while LQ dominates in advection-dominated regimes. The ratio $\Delta D_{\mathrm{LQ}}/\Delta D_{\mathrm{TQ}}$ exhibits a smooth inverse-square dependence on the dynamo effectivity range, refining previous empirical fits with improved accuracy and reduced scatter. The results further reveal that the need for a decay term is not essential for PINN setup due to the training process. Compared with the traditional 1D SFT model, the PINN framework achieves significantly lower error metrics and more robust recovery of nonlinear trends. Our results suggest that the nonlinear interplay between LQ and TQ can naturally produce alternations between weak and strong cycles, providing a physical explanation for the observed even–odd cycle modulation. These findings demonstrate the potential of PINN as an accurate, efficient, and physically consistent tool for solar cycle prediction.

## 1. Introduction

The solar dynamo is a fundamental process responsible for maintaining the Sun's magnetic field (E. N. Parker 1955, 1958). This mechanism operates within the Sun's interior, particularly in the convection zone, where the interplay of turbulent plasma motions and rotational effects creates and sustains magnetic fields through magnetohydrodynamic processes (S. M. Tobias 2009; P. Charbonneau 2014; T. Sekii & H. Shibahashi 2015; P. Charbonneau & D. Sokoloff 2023). The solar cycle, approximately 11 yr long, is characterized by the periodic increase and decrease in the number of sunspots. This cycle is driven by the solar dynamo, which modulates the magnetic field's strength and polarity over time (P. Charbonneau 2010; J. A. Abreu et al. 2012; P. Bhowmik et al. 2023). Sunspots are temporary activity on the Sun's photosphere that appear as spots darker than the surrounding areas. They are regions of intense magnetic activity, and their cyclic appearance is a direct manifestation of the solar dynamo's operation (P. Charbonneau 2014; P. Bhowmik et al. 2023; G. M. Vasil et al. 2024). The Sun's equator rotates faster than its poles, creating shear that stretches and twists magnetic field lines, contributing to the generation of the toroidal magnetic field component (R. K. Ulrich & J. E. Boyden 2005; A. R. Choudhuri 2023). This large-scale meridional circulation, together with differential rotation, transports magnetic flux on the solar surface from low latitudes toward the poles and plays a significant role in regulating the solar cycle dynamics (H. Babcock 1961; R. B. Leighton 1969; A. R. Choudhuri 2021). The $\alpha$-effect involves the twisting of magnetic field lines due to helical turbulence, while the $\omega$-effect refers to the shearing of poloidal magnetic fields into toroidal fields by differential rotation (E. N. Parker 1955; M. Steenbeck 1966).

The surface flux transport (SFT) model plays a crucial role in simulating the evolution of solar magnetic fields by describing the transport and redistribution of magnetic flux on the Sun's surface (R. B. Leighton 1964; Y.-M. Wang et al. 1989; A. R. Yeates et al. 2023 and references therein). It incorporates various mechanisms such as differential rotation, meridional circulation, and supergranular diffusion to describe the transport and evolution of magnetic flux (R. H. Cameron et al. 2012; J. Jiang et al. 2014b). The model starts with the emergence of magnetic bipoles and uses the induction equation to track the evolution of the magnetic field, assuming it behaves as if it were purely radial after emergence (R. B. Leighton 1964; J. Jiang et al. 2014b).

The formation of the toroidal field from the poloidal field is a linear process. In contrast, the transformation of the toroidal field back into a poloidal field is nonlinear, governed by mechanisms such as tilt quenching (TQ; M. Dasi-Espuig et al. 2010; A. Lemerle & P. Charbonneau 2017; B. B. Karak & M. S. Miesch 2017; B. B. Karak & M. Miesch 2018; B. K. Jha et al. 2020) and latitude quenching (LQ; K. J. Li et al. 2003; S. K. Solanki et al. 2008; A. G. Tlatov & A. A. Pevtsov 2010; J. Jiang et al. 2011; J. Jiang 2020), which act on the source term implemented within the SFT framework (P. Charbonneau et al. 2005; K. Petrovay 2020; M. Talafha et al. 2022). TQ refers to the tendency of stronger cycles to produce active regions (ARs) with smaller tilts relative to the equator (B. K. Jha et al. 2023), while LQ corresponds to the emergence of ARs at higher latitudes during stronger cycles (A. R. Yeates et al. 2025).

These two mechanisms are not the only ones to consider; other attempts have been made to include surface inflows toward ARs as a nonlinearity to the model (D. Martin-Belda & R. H. Cameron 2017; M. Nagy et al. 2017; K. Teweldebirhan et al. 2024; M. H. Talafha et al. 2025). All of the possible nonlinearities come with a great value in completing the poloidal-toroidal formation. Nonlinearities like TQ and LQ are essential for the predictability and stability of the solar magnetic activity cycle. They ensure that the solar cycle amplitude remains within a limited range, preventing extreme variations that could disrupt the solar dynamo process (L. Deng et al. 2015; J. Jiang 2020). The nonlinear feedback mechanisms provided by TQ and LQ are critical in the Babcock–Leighton dynamo model. These mechanisms help explain observed long-term solar cycle variability, such as the Gnevyshev–Ohl rule, which describes the alternation of strong and weak solar cycles (A. R. Yeates et al. 2025).

Expanding on the methodology of (J. Jiang et al. 2014a), who investigated the asymptotic dipole contribution of ARs depending on their properties (K. Petrovay et al. 2020) introduced the dynamo effectivity range ($\lambda_R$), which quantifies the latitudinal extent over which the surface flows efficiently transport magnetic flux before it is dissipated by diffusion. Dynamo effectivity range plays a crucial role in determining the relative effectiveness of TQ to LQ, and consequently, $\lambda_R$ provides a natural scaling parameter for comparing and quantifying the contributions of these nonlinear feedback mechanisms across different dynamo models.

Traditional grid-based numerical models and analytical approaches face several limitations when addressing the nonlinear processes inherent in the solar dynamo. This includes complex nonlinear interactions, such as the coupling between polar magnetic fields and AR fields (L. Deng et al. 2015; Z. Shuang et al. 2015).

Physics-informed neural networks (PINNs) have emerged as a significant advancement in the intersection of machine learning and physical sciences, offering innovative solutions for complex problems governed by physical laws. PINNs incorporate physical laws, often described by partial differential equations (PDEs), directly into the neural network training process. This integration helps in achieving high accuracy even with limited data (S. Falas et al. 2020; S. Markidis 2021; B. Wang et al. 2023). It is known for its ability to solve problems with minimal data, leveraging the underlying physics to guide the learning process. This results in efficient and accurate modeling of physical systems (S. Falas et al. 2020; C. Lenz et al. 2024; Y. Song et al. 2024). PINNs demonstrate robustness in scenarios with sparse or noisy data by incorporating physical constraints, which helps in maintaining predictive accuracy (M. Yang & J. T. Foster 2022; T. Li et al. 2024; Z. Zou et al. 2024). PINNs are used to solve various PDEs in scientific computing, including fluid dynamics, structural analysis, and turbulence modeling (S. Markidis 2021; W. Fox et al. 2024; S. Khalid et al. 2024). They are employed in modeling complex systems where traditional methods may fall short, such as in the simulation of localized waves and nonlinear Schrödinger equations (Y. Guo et al. 2023)

PINNs offer a promising alternative framework for solving the solar SFT problem, particularly in capturing its nonlinear behavior and enhancing predictive capability. By embedding the governing SFT equations directly into the neural network's loss function, PINNs can learn solutions consistent with both physical laws and available magnetic field data. Their mesh-independent formulation ensures better numerical stability and flux conservation, which are crucial for accurately reproducing observed polar field evolution and enhancing solar cycle predictions. While the present work focuses on the physics-constrained implementation, further extensions could incorporate data assimilation and multiparameter optimization to refine the representation of nonlinear effects and improve forecasting accuracy (S. Moschou et al. 2023; J. J. Athalathil et al. 2024; M. Gholampour et al. 2024; J. Zhang et al. 2024).

This work investigates the impact of incorporating nonlinearities of the solar cycle modulation as TQ and LQ in the SFT model using PINN techniques and compares the result with the literature. We further aim to understand the complex interplay between these quenching mechanisms in the buildup of the polar field from one cycle to the next. This article is structured as follows: Section 2 describes the SFT model and implementation using PINNs, highlighting the source model and nonlinearities adopted in this study. In Section 3, we explain how including the nonlinearities affected the model. Section 4 compares the results of PINNs and a simple upwind scheme used in (K. Petrovay & M. Talafha 2019) and (M. Talafha et al. 2022) to solve the SFT equation. Section 5 presents an interpretation from this study on the variation of the solar cycle, and Section 6 concludes the study.

## 2. SFT-PINN Framework

### 2.1. SFT Model

In this work, we simulate a 1D SFT model within a Babcock–Leighton-type dynamo framework, as described in (M. Talafha et al. 2022). SFT equation (Equation (1)) is a transport equation for the radial magnetic field derived from the induction equation. The decay term was included later to address the problem of abrupt drift in the polar field over time. This issue arises when fluctuations in cycle amplitude lead to a random walk of the unsigned solar dipole moment from one cycle to the next, eventually reaching values far exceeding the typical total contribution of ARs in a cycle (C. J. Schrijver et al. 2002)

$$\frac{\partial B}{\partial t} = \frac{1}{R\cos\lambda}\frac{\partial}{\partial\lambda}(B\,u\cos\lambda) + \frac{\eta}{R^2\cos\lambda}\frac{\partial}{\partial\lambda}\left(\cos\lambda\frac{\partial B}{\partial\lambda}\right) - \frac{B}{\tau} + S(\lambda, t), \qquad (1)$$

where $R$ is the solar radius, $u$ is the meridional flow, $\eta$ is the diffusivity, and $\tau$ represents the decay time. The source term $S(\lambda, t)$ represents the parameterized source function describing the spatiotemporal emergence of p- and f-polarities on the solar surface as described in (K. Petrovay & M. Talafha 2019). In their work, the source is represented as a pair of rings of opposite magnetic polarity as

$$\begin{aligned} S(\lambda, t) = & kA_m f_{\rm sph} S_1(t) S_2[\lambda;\ \lambda_0(t) - \Delta\lambda(t),\ \delta\lambda] \\ & - kA_m S_1(t) S_2[\lambda;\ \lambda_0(t) + \Delta\lambda(t),\ \delta\lambda] \\ & + kA_m S_1(t) S_2[\lambda;\ -\lambda_0(t) - \Delta\lambda(t),\ \delta\lambda] \\ & - kA_m f_{\rm sph} S_1(t) S_2[\lambda;\ -\lambda_0(t) + \Delta\lambda(t),\ \delta\lambda], \end{aligned} \quad (2)$$

where $k = \pm 1$ is a factor alternating between even and odd cycles. $A_m$ is an arbitrary amplitude depending on the flow profile (0.015 for a simple sinusoidal profile) used to ensure that the resulting polar field amplitude roughly agrees with observations. Care was taken to ensure zero net flux on the spherical surface by reducing the amplitude of the equatorward member of each pair by an appropriate sphericity factor $f_{\rm sph}$.

$S_1(t)$ is the time profile of solar activity in a typical cycle was determined by D. H. Hathaway et al. (1994) from the average of many cycles as

$$S_1(t) = a t_c^3 / [\exp(t_c^2 / b^2) - c], \quad (3)$$

with $a = 0.00185$, $b = 48.7$, $c = 0.71$, where $t_c$ is the time since the last cycle minimum.

$S_2[\lambda;\ \lambda_0(t),\ \delta\lambda]$ is the latitudinal profile represented by a Gaussian with a fixed full width at half-maximum of $2\delta\lambda = 6\overset{\circ}{.}26$, migrating equatorward during the course of a cycle:

$$S_2(\lambda;\ \lambda_0,\ \delta\lambda) = \frac{\delta\lambda_0}{\delta\lambda} \exp[-(\lambda - \lambda_0)^2 / 2\delta\lambda^2], \quad (4)$$

while its standard deviation ($\delta\lambda$) follow the empirical results of J. Jiang et al. (2011):

$$\delta\lambda = [0.14 + 1.05(t/P) - 0.78(t/P)^2]\lambda_0, \quad (5)$$

where $P$ is the cycle period (11 yr). The latitudinal separation of the rings is a consequence of Joy's law:

$$2\Delta\lambda = 0.5 \frac{\sin\lambda}{\sin 20^\circ}, \quad (6)$$

while the mean latitude $\lambda_0$ of activity during the course of a cycle at a given phase of cycle i, is given by a quadratic fit derived by J. Jiang et al. (2011) from many observed solar cycles:

$$\lambda_0(t,\ i) = 26.4 - 34.2(t/P) + 16.1(t/P)^2, \quad (7)$$

and if it is parameterized as a function of cycle strength, it represents the LQ nonlinearity (M. Talafha et al. 2022):

$$\begin{aligned} \lambda_0(t,\ i) = & [26.4 - 34.2(t/P) + 16.1(t/P)^2] \\ & \times \left(1 + \frac{b_{\rm lat}}{14.6}\left(\frac{A_n}{A_0} - 1\right)\right), \end{aligned} \quad (8)$$

With $b_{\rm lat}$ determining the rate at which the emergence latitude increases and is set to 2.4 based on the work in J. Jiang et al. (2011), $(A_n/A_0)$ denotes the strength of a given cycle relative to a reference cycle. We model fluctuations as multiplicative noise acting on the amplitude of the poloidal field source. Hence, the cycle amplitude follows a lognormal distribution described as $A_n = A_0 \times 10^G$, where $G$ is a Gaussian random variable with zero mean and standard deviation 0.13. The normalization constant is set to $A_0 = 0.001\, e^{7/\tau}$ (Gauss), which yields dipole moment values comparable to observations.

TQ is introduced by multiplying the nominal Joy's law tilt with a suppression factor that scales with cycle amplitude:

$$\Delta\lambda = \delta_0 \sin\lambda_0 \left(1 - b_{\rm joy}\left(\frac{A_n}{A_0} - 1\right)\right), \quad (9)$$

where $\delta_0 = 1.5$ is Joy's law amplitude coefficient, and the constant coefficient $b_{\rm joy} = 0.15$ specifies the quenching efficiency. Together, TQ and LQ act to suppress the buildup of the large-scale dipole. While TQ limits the separation of opposite polarities within emerging ARs, LQ reduces the efficiency of flux cancellation across the equator.

According to K. Petrovay et al. (2020), the dynamo effectivity range, $\lambda_R$, characterizes the competition between advective transport by the meridional flow and diffusive transport due to turbulent mixing in the surface magnetic flux evolution. It is defined as

$$\lambda_{\rm R} = \sqrt{\frac{\eta}{u_0 R}}, \quad (10)$$

where $u_0$ is the characteristic meridional flow amplitude. This parameter plays a central role in determining the relative impact of LQ and TQ on the cycle-integrated dipole moment, as discussed in M. Talafha et al. (2022). In particular, the ratio of their respective deviation scales as

$$\Delta_{LQ}/\Delta_{TQ} \sim C_1(\bar{\lambda_0}) + C_2(\bar{\lambda_0})/\lambda_R^2, \quad (11)$$

where $\Delta_{\rm LQ}$ and $\Delta_{\rm TQ}$ denote the deviations in the net contribution of a cycle to the dipole moment from the no quenching case, arising from LQ and TQ, respectively.

The net contribution of each cycle to the buildup of the large-scale dipole is defined as the residual dipole moment:

$$D_{\rm res}(P,\ \tau) = D(t) - e^{-P/\tau} D_{\rm cycle\ min}, \quad (12)$$

where $D(t)$ is the instantaneous dipole moment, $D_{\rm cycle\ min}$ is the value at the preceding cycle minimum. This formulation removes the exponential memory of earlier cycles, allowing us to quantify the effective dipole amplification within a single cycle. We use $D_{\rm res}(P,\ \tau)$ as a diagnostic throughout this study to compare the influence of TQ versus LQ.

### 2.2. PINNs

The input layer of the PINN model consists of latitude ($\lambda$) and time ($t$), while the output layer predicts the magnetic field $B(\lambda,\ t)$. A brief description of the model is provided below, with a detailed explanation available in our previous work (J. J. Athalathil et al. 2024).

For each training point ($\lambda$, $t$), the network predicts the magnetic field $B_{\rm pinn}(\lambda,\ t)$. The derivatives required by the SFT equation are computed using automatic differentiation. Since the network outputs are linked to the inputs through activation functions with known analytical derivatives, derivatives of $B_{\rm pinn}(\lambda,\ t)$ with respect to ($\lambda$, $t$) can be evaluated directly using the chain rule. This grid-free approach enables derivative evaluation at any point within the domain and avoids explicit discretization of both spatial and temporal coordinates.

The loss function is constructed by enforcing the governing SFT equation together with the prescribed initial and boundary conditions at the training points. The neural network weights are then updated via backpropagation to minimize the loss. This iterative optimization continues until the loss converges. We employ a two-step optimization strategy to train the model. Initially, the ADAM optimizer (D. P. Kingma & J. Ba 2014) is used to achieve rapid convergence in the early stages of training. Subsequently, the L-BFGS optimizer (D. C. Liu & J. Nocedal 1989) is applied to fine-tune the solution and achieve higher precision in the final optimization.

The loss function for the learning process is constructed as a weighted sum of three components: initial condition loss, boundary condition loss, and PDE residual loss. These components collectively ensure that the PINNs solution adheres to the specified initial conditions, satisfies the prescribed boundary conditions, and minimizes the residual of the governing PDE. Mathematically, the loss function is expressed as

$$\xi = w_{\rm ic}\xi_{\rm ic} + w_{\rm bc}\xi_{\rm bc} + w_{\rm pde}\xi_{\rm pde}. \tag{13}$$

Here, the weights $w_{\rm ic}$, $w_{\rm bc}$, and $w_{\rm pde}$ balance the contributions of each component to the total loss. This formulation ensures that the model learns a solution consistent with both the physical constraints and the observed data. The components of the loss function $\xi_{\rm ic}$, $\xi_{\rm bc,}$ and $\xi_{\rm pde}$ correspond to the initial condition loss, boundary condition loss, and PDE residual loss, respectively. These components are mathematically defined as

$$\xi_{\rm ic} = \frac{1}{N_{\rm ic}}\sum_{\rm i=1}^{N_{\rm ic}} |B_{\rm pinn}(\lambda^{\rm i}, t=0) - B_{\rm init}(\lambda^{\rm i})|^2, \tag{14}$$

$$\xi_{\rm bc} = \frac{1}{N_{\rm bc}}\sum_{\rm j=1}^{N_{\rm bc}} |B_{\rm pinn}(\lambda_{\rm bc}^{\rm j}, t^{\rm j}) - B_{\rm bc}(\lambda_{\rm bc}^{\rm j}, t^{\rm j})|^2, \tag{15}$$

$$\xi_{\rm pde} = \frac{1}{N_{\rm pde}}\sum_{\rm k=1}^{N_{\rm pde}} |f(\lambda^{\rm k}, t^{\rm k})|^2, \tag{16}$$

where $f(\lambda, t)$ is the residual function that represents the error in satisfying the SFT equation as given below:

$$f(\lambda, t) = \frac{\partial B_{\rm pinn}}{\partial t} - \frac{1}{R\cos\lambda}\frac{\partial}{\partial\lambda}(B_{\rm pinn}u(\lambda)\cos(\lambda)) - \frac{\eta}{R^2\cos\lambda}\frac{\partial}{\partial\lambda}\left(\cos\lambda\frac{\partial B_{\rm pinn}}{\partial\lambda}\right) + \frac{B_{\rm pinn}}{\tau} - S(\lambda, t). \tag{17}$$

Here, $B_{\rm pinn}(\lambda, t)$ represents the predicted magnetic field at each point $(\lambda, t)$. The terms $N_{\rm ic}$, $N_{\rm bc}$, and $N_{\rm pde}$ denote the number of training points used for the initial condition, boundary condition, and domain residual calculations, respectively. $B_{\rm init}(\lambda)$ is the initial condition (at $t = 0$), while $B_{\rm bc}(\lambda_{\rm bc}, t)$ represents the boundary conditions at boundary $\lambda_{\rm bc}$. The initial condition is taken to be a dipolar field $B_{\rm init}(\lambda) \equiv \sin\lambda$. Due to the presence of the decay term in the SFT equation, this dipolar initial magnetic field decays within a few years. To ensure that the choice of initial condition does not influence the final results, we discard the first two solar cycles (22 yr) from all simulations and base our analysis only on the subsequent evolution. Neumann boundary conditions with zero gradient are imposed at the latitudinal boundaries, allowing unimpeded poleward transport of magnetic flux consistent with the SFT model.

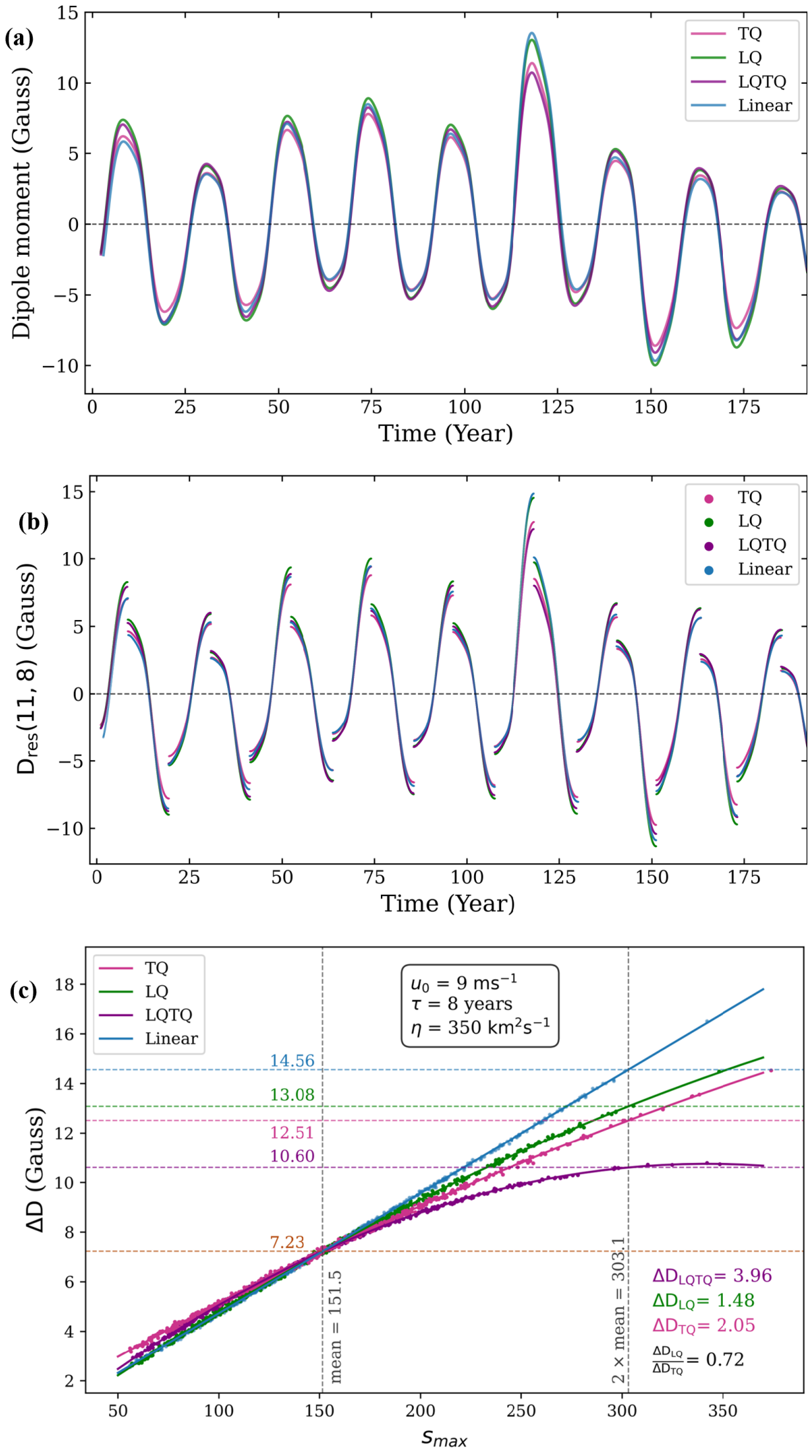

**Figure 1.** Simulation results for SFT parameters $u_0 = 9$ m s$^{-1}$, $\tau = 8$ yr, and $\eta = 350$ km$^2$ s$^{-1}$. (a) Dipole moment distribution in Gauss across latitudes for each time step over 15 cycles. (b) Time evolution of the residual dipole moment,$D_{\rm res}(P = 11, \tau = 8)$ (c) Change in dipole moment of each cycle from the previous cycle for different processes (TQ, LQ, LQ+TQ, and linear). Deviations are quantified at twice the mean of the maximum source.

## 3. Results

A comprehensive parameter study was carried out by systematically varying the SFT parameters through four different cases: No quenching case (linear), TQ, LQ, and the combination of both LQ and TQ (LQTQ). The results corresponding to $u_0 = 9$ m s$^{-1}$, $\tau = 8$ yr, and $\eta = 350$ km$^2$s$^{-1}$ are presented as an example.

Figure 1(a) shows the time evolution of the axial dipole moment in Gauss. During weaker cycles, all models exhibit similar dipole amplitudes. However, during the strong cycle, TQ reduces the mean Joy's law tilt, increasing cancellation between opposite polarities within a hemisphere, while LQ shifts flux emergence poleward, reducing cross-equatorial transport. Acting together, these effects substantially suppress the net polar dipole buildup, producing the prominent deviation seen around year ∼120. The combined LQTQ model consistently yields the lowest dipole amplitudes, reflecting the combined regulatory effects of both feedback mechanisms. The overall magnetic dipole moment shown in Figure 1(a) also has a contribution from the preceding cycles. These cumulative contributions from prior cycles need to be removed to accurately capture the net buildup of the dipole moment within each solar cycle. Figure 1(b) shows the time evolution of the residual dipole moment $D_{\rm res}$, as defined in Equation (12). This diagnostic isolates cycle-to-cycle amplification and reveals Gaussian-like growth curves for each solar cycle, with their peaks indicating the rate of dipole amplification. The plot shows that while all simulated SFT models exhibit similar overall temporal patterns, subtle differences arise in the amplitude and shape of individual cycles due to the influence of the quenching mechanisms.

The gain of the dipole moment per cycle represents the net accumulation of the polar magnetic field throughout a cycle. We calculated the gain (ΔD) as the residual dipole moment at the end of each cycle ($n$) at the mean and twice the mean of the maximum source amplitude injected during the cycle ($S_{\rm max}$, $n$), which serves as a ballpark for the strength of the given cycle. The variation of ΔD as a function of the $S_{\rm max}$, $n$ of the cycle is shown in Figure 1(c). The linear case is used as a reference to understand the effect of quenching. In the linear regime, ΔD (and hence the dipole moment gain) increases proportionally with the source amplitude. The inclusion of TQ and LQ introduces significant nonlinear saturation, particularly at higher source strengths. The combined effect (LQTQ) shows the strongest suppression, resulting in the lowest ΔD values. For lower amplitude cycles, the nonlinear cases can produce slightly enhanced dipole moment gains relative to the linear case.

We calculate the deviations due to different quenching mechanisms ($\Delta D_{\rm LQ}$, $\Delta D_{\rm TQ}$, $\Delta D_{\rm LQTQ}$) as the change in $\Delta D$ evaluated from the linear case at twice the mean of $S_{{\rm max},n}$. The deviation in dipole moment due to the combined nonlinearities ($\Delta D_{\rm LQTQ} = 3.96$) is larger than that from either mechanism alone ($\Delta D_{\rm TQ} = 2.04$ for TQ and $\Delta D_{\rm LQ} = 1.48$ for LQ). The resulting ratio $\Delta D_{\rm LQ}/\Delta D_{\rm TQ} = 0.72$ indicates that TQ is the more effective mechanism in suppressing the polar dipole moment, while LQ provides a secondary but reinforcing contribution for the chosen set of SFT parameters. These results complement previous findings by M. Talafha et al. (2022).

Figure 2 presents the parameter sensitivity of the nonlinear quenching effects, showing the variations of $\Delta D_{\rm LQ}$, $\Delta D_{\rm TQ}$, and their ratio $\Delta D_{\rm LQ}/\Delta D_{\rm TQ}$ across the SFT parameter space. Panels (a)–(f) show the two-dimensional sensitivity maps of $\Delta D_{\rm LQ}$, $\Delta D_{\rm TQ}$, and their ratio in the ($u_0$, $\eta$) parameter space for $\tau = 8$ yr (left column) and $\tau = \infty$ (right column). Warmer colors (dark orange to red) represent larger deviations of $\Delta D_{\rm LQ}$, indicating stronger suppression of the dipole moment in diffusion-dominated regimes, whereas cooler tones (blue to cyan) correspond to weaker suppression. In contrast, $\Delta D_{\rm TQ}$ varies less systematically but remains more pronounced in advection-dominated regimes (high $u_0$, low $\eta$). The ratio map reveals that $\Delta D_{\rm LQ}/\Delta D_{\rm TQ}$ is largest at low diffusivity and high flow speed, gradually declining as $\eta$ increases. Together, these results confirm that LQ dominates in advection-dominated regions, whereas TQ becomes increasingly effective in diffusion-dominated regimes. The $\Delta D_{\rm LQ}$ and $\Delta D_{\rm TQ}$ for different SFT parameters are presented in Table 1.

A comparative analysis of the ratio $\Delta D_{\rm LQ}/\Delta D_{\rm TQ}$ as a function of $\lambda_R$ is shown in Figure 3 for two decay timescales: $\tau = 8$ [yr] and $\tau = \infty$, respectively. The green and magenta shaded regions mark the LQ-dominated and TQ-dominated regimes, respectively. In both panels, the teal points correspond to the PINN results, while the crimson markers show the values from M. Talafha et al. (2022). With increasing $\lambda_R$, LQ becomes progressively less influential compared to TQ. This behavior is particularly evident at higher values of $\lambda_R$, where the contribution of TQ increasingly dominates the modulation of the dipole moment, and in turn, the strength of the next cycle. A nonlinear fit of the form $C_1 + C_2/\lambda_R^2$ is applied to the PINN data, yielding the best fit (indicated by the dashed black curve in Figure 3). The best-fit values are found to be $C_1 = -0.3847$ and $C_2 = 136.8595$ ($\tau = 8$ yr). For a comparison of the fitted parameters and a detailed mathematical interpretation of the corresponding formula, see M. H. Talafha (2025). The PINN data exhibit a smooth and consistent trend that closely follows the expected inverse-square dependence as compared to the values in M. Talafha et al. (2022). A detailed comparison is presented in Section 4, highlighting the improved reliability and reduced variance in the PINN-based results, reinforcing the accuracy and robustness of the PINN framework in capturing the underlying physical behavior of the system.

## 4. Improved Accuracy with PINNs

The improved agreement of the PINN model with the expected inverse-square behavior, as observed in Figure 3 (a) and (b), motivates a deeper evaluation of its quantitative performance. Table 2 presents a comparative analysis of the dipole moment predictions from our PINN-based approach and the numerical scheme by M. Talafha et al. (2022) using standard performance metrics. The comparison is performed between the model-predicted $\Delta D_{\rm LQ}/\Delta D_{\rm TQ}$ values and their corresponding fitted profiles to assess the scatter and to evaluate how well the underlying nonlinear trend is captured. The definitions of the error metrics, along with their distributions, are provided in Appendix A. The mean absolute error (MAE), mean squared error (MSE), and root mean squared error (RMSE) values are significantly lower for the PINN model, indicating better overall predictive accuracy. The distribution for these error metrics has low scatter for PINNs (see Figure 5) thus indicating improved model performance.

With a coefficient of determination ($R^2$) score of 0.9841, the PINN model exhibits an even closer match to the best-fit profile, improving upon the 0.9730 achieved in M. Talafha et al. (2022). The maximum error and mean absolute percentage error (MAPE) are also considerably reduced in the PINN case, highlighting the model's robustness across the entire range of inputs. These results suggest that the PINN framework offers improved accuracy and reliability for modeling nonlinear quenching effects. Unlike traditional schemes that may suffer from numerical sensitivity to the resolution of the simulation box, PINNs naturally incorporate physical constraints and exhibit greater stability across the dynamo parameter space.

In terms of computational efficiency, each PINN run (training on a single set of SFT parameters) typically required ∼1–2 hr on a standard CPU node (Intel i7, 4 cores), or about 15–20 minutes on a single GPU (NVIDIA V100). In comparison, the traditional upwind finite-difference scheme completes within a few minutes on the same CPU, but suffers from greater numerical scatter and resolution sensitivity. Thus, while the PINN incurs a higher per-run training cost, it yields smoother and more accurate solutions across the parameter space, particularly in the nonlinear regimes of interest.

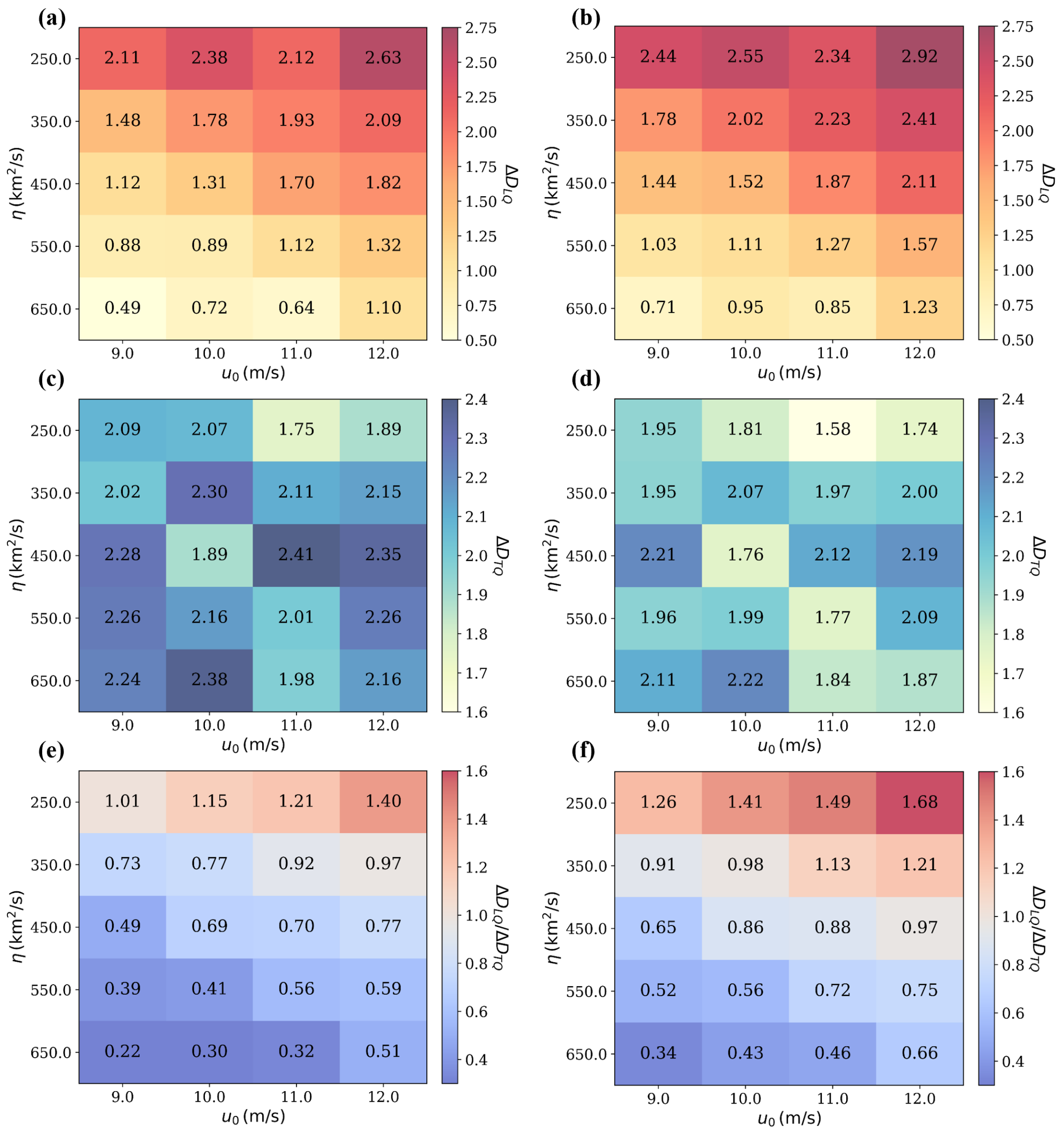

**Figure 2.** Parameter sensitivity of the nonlinear quenching effects across the SFT parameter space. Panels (a), (c), (e) and (b), (d), (f) show the sensitivity maps of $\Delta D_{\rm LQ}$, $\Delta D_{\rm TQ}$, and their ratio across the ($u_0$, $\eta$) parameter space for $\tau = 8$ and $\tau = \infty$, respectively.

Although retuning the hyperparameters is not necessary in the PINN architecture, the model must be retrained whenever the initial conditions or simulation parameters are altered (J. J. Athalathil et al. 2024). In the context of the present work, retraining is required each time the SFT parameters ($u_0$, $\eta$, and $\tau$) are modified. This repeated retraining increases the overall computational cost and partially offsets the efficiency gains expected from machine learning approaches. Recent developments, such as neural operators including the Fourier neural operator and its physics-informed variants (Z. Li et al. 2020; N. Kovachki et al. 2021; Z. Li et al. 2021; W. Zhong & H. Meidani 2025) have shown potential in addressing this retraining bottleneck. Similarly, reduced-order surrogate modeling approaches (P. Conti et al. 2024) and parameter conditioning strategies (Y. Chen et al. 2022; V. Madhavan et al. 2024) aim to enhance adaptability across parameter spaces. Together, these methods offer promising routes toward minimizing the need for repeated training and improving computational efficiency.

## 5. Discussion

Through this work, we explored the SFT parameter space to understand the role of quenching mechanisms in dipole moment buildup in a Babcock–Leighton-type solar dynamo. Using the PINN framework, we independently verified the inverse relationship between dipole moment gain and the dynamo effectivity range $\lambda_R$, previously reported in M. Talafha et al. (2022). The ratio $\Delta D_{\rm LQ}/\Delta D_{\rm TQ}$ derived from a PINN follows a smooth inverse-square trend, accurately captured by the empirical relation $C_1 + C_2/\lambda_R^2$, as shown in Figure 3.

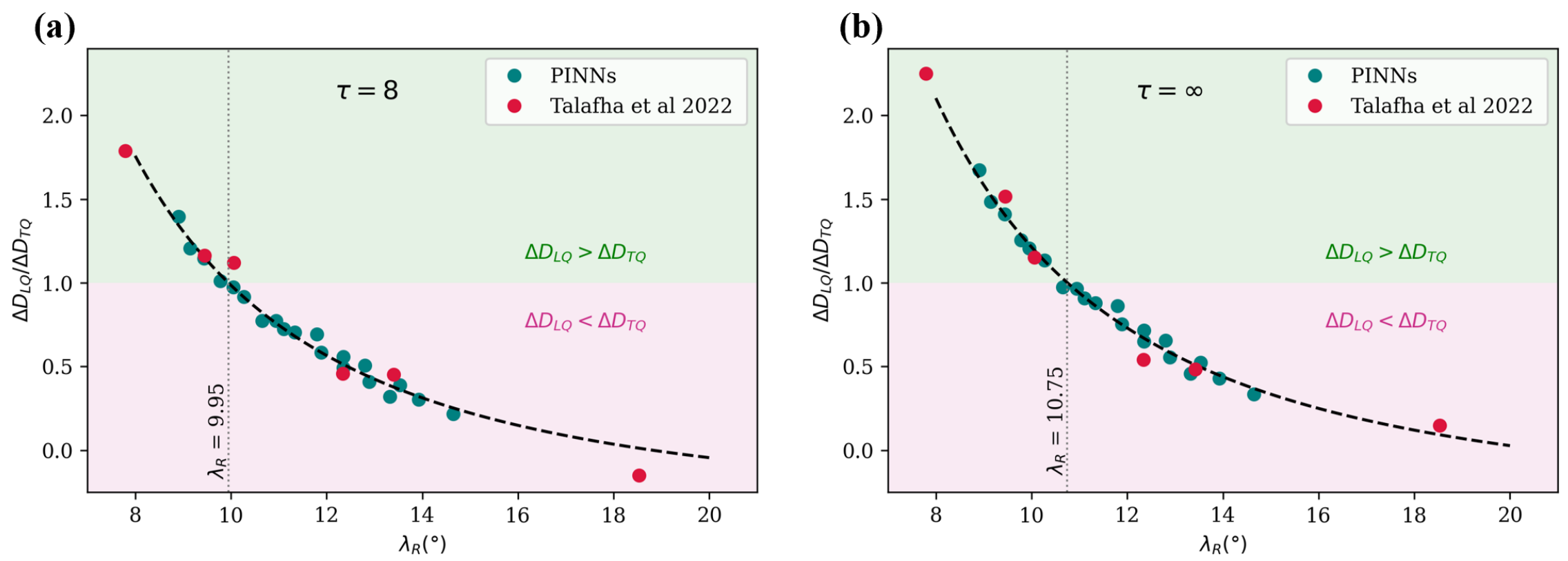

**Figure 3.** Relative importance of nonlinear quenching mechanisms. Ratio $\Delta D_{\rm LQ}/\Delta D_{\rm TQ}$ is plotted as a function of the dynamo effectivity range $\lambda_R$ for finite $\tau = 8$ (a) and $\tau = \infty$ (b). The teal symbols show the PINN results, the dashed black curve is the best-fit inverse-square law ($C_1 + C_2/\lambda_R^2$), and the crimson points are values from M. Talafha et al. (2022). Best-fit parameters: $C_1 = -0.3847$, $C_2 = 136.86$ ($\tau = 8$); $C_1 = -0.3679$, $C_2 = 158.19$ ($\tau = \infty$). The shaded regions highlight regimes dominated by LQ (green) and TQ (magenta), showing that LQ dominates at low $\lambda_R$ (advection-dominated regimes) and TQ becomes increasingly effective at higher $\lambda_R$ (diffusion-dominated regimes).

**Table 1**
Comparison of Dipole Moment Deviations $\Delta D_{\rm LQ}$ and $\Delta D_{\rm TQ}$ for Different Values of $u_0$ and $\eta$

| $u_0$ (m s$^{-1}$) | $\eta$ (km$^2$ s$^{-1}$) | $\lambda_R$ (deg) | $\tau = 8$ | | | $\tau = \infty$ | | |
|---|---|---|---|---|---|---|---|---|
| | | | $\Delta D_{\rm LQ}$ | $\Delta D_{\rm TQ}$ | $\Delta D_{\rm LQ}/\Delta D_{\rm TQ}$ | $\Delta D_{\rm LQ}$ | $\Delta D_{\rm TQ}$ | $\Delta D_{\rm LQ}/\Delta D_{\rm TQ}$ |
| 9 | 250 | 9.78 | 2.11 | 2.09 | **1.01** | 2.44 | 1.95 | 1.25 |
| 10 | 250 | 9.44 | 2.38 | 2.07 | 1.15 | 2.55 | 1.81 | 1.41 |
| 11 | 250 | 9.15 | 2.12 | 1.75 | 1.21 | 2.34 | 1.57 | 1.49 |
| 12 | 250 | 8.91 | 2.63 | 1.89 | 1.40 | 2.92 | 1.74 | 1.68 |
| 9 | 350 | 11.10 | 1.48 | 2.04 | 0.72 | 1.78 | 1.95 | 0.91 |
| 10 | 350 | 10.65 | 1.78 | 2.30 | 0.77 | 2.02 | 2.07 | **0.98** |
| 11 | 350 | 10.27 | 1.93 | 2.11 | 0.92 | 2.23 | 1.97 | 1.13 |
| 12 | 350 | 9.95 | 2.09 | 2.15 | **0.97** | 2.41 | 2.00 | 1.21 |
| 9 | 450 | 12.35 | 1.12 | 2.28 | 0.49 | 1.44 | 2.21 | 0.65 |
| 10 | 450 | 11.79 | 1.31 | 1.89 | 0.69 | 1.52 | 1.76 | 0.86 |
| 11 | 450 | 11.33 | 1.70 | 2.41 | 0.70 | 1.87 | 2.12 | 0.88 |
| 12 | 450 | 10.94 | 1.82 | 2.35 | 0.77 | 2.11 | 2.19 | **0.97** |
| 9 | 550 | 13.53 | 0.88 | 2.26 | 0.39 | **1.03** | 1.96 | 0.52 |
| 10 | 550 | 12.89 | 0.89 | 2.16 | 0.41 | 1.11 | 1.99 | 0.56 |
| 11 | 550 | 12.35 | 1.12 | 2.00 | 0.56 | 1.27 | 1.76 | 0.72 |
| 12 | 550 | 11.89 | 1.32 | 2.26 | 0.58 | 1.57 | 2.09 | 0.75 |
| 9 | 650 | 14.65 | 0.48 | 2.24 | 0.22 | 0.71 | 2.11 | 0.34 |
| 10 | 650 | 13.93 | 0.72 | 2.38 | 0.30 | **0.95** | 2.22 | 0.43 |
| 11 | 650 | 13.32 | 0.64 | 1.98 | 0.32 | 0.84 | 1.84 | 0.46 |
| 12 | 650 | 12.80 | 1.10 | 2.16 | 0.51 | 1.23 | 1.87 | 0.65 |

**Note.** The ratio quantifies the relative importance of the two quenching mechanisms for finite $\tau = 8$ and $\tau \rightarrow \infty$. Transition values ($\Delta D_{\rm LQ}/\Delta D_{\rm TQ} \approx 1$) are shown in bold.

**Table 2**
Comparison of Error and Fit Metrics between the PINN Model and Those of M. Talafha et al. (2022) for $\tau = 8$

| Metric | PINN | M. Talafha et al. (2022) |
|---|---|---|
| MAE | 0.032293 | 0.090180 |
| MSE | 0.001588 | 0.010587 |
| RMSE | 0.039855 | 0.102894 |
| $R^2$ | 0.984119 | 0.972994 |
| Max error | 0.093414 | 0.158281 |
| MAPE (%) | 6.017877 | 23.306108 |

**Note.** Details of the metric calculation are given in Appendix A

The PINN framework successfully recovers the observed trend and further refines the empirical fit by enforcing physical consistency, leading to tighter constraints on the governing coefficients. This establishes the PINN as a powerful tool for both prediction and insight in solar dynamo modeling. In our subsequent paper, we will use this model to predict the next solar cycle by training the neural network on Wilcox Solar Observatory (WSO) synoptic maps.

We found that the efficiency of nonlinear quenching mechanisms is inherently dependent on the underlying transport regime. The ratio $\Delta D_{\rm LQ}/\Delta D_{\rm TQ}$ serves as a diagnostic indicator of this behavior, and its variation with $\lambda_R$ reveals a distinct transition in dominance between the two mechanisms. Particularly, the point at which the ratio crosses unity (i.e., $\Delta D_{\rm LQ}/\Delta D_{\rm TQ} = 1$) represents a critical threshold beyond which TQ becomes the dominant nonlinear feedback, replacing LQ. This transition typically occurs around $\lambda_R \approx 10^\circ$ ($9^\circ\!\!.95$ for $\tau = 8$ and $10^\circ\!\!.75$ for $\tau = \infty$), suggesting that LQ is more effective in advection-dominated regimes (low $\lambda_R$), whereas TQ becomes increasingly significant in diffusion-dominated regimes (high $\lambda_R$). This agrees with the findings of M. H. Talafha et al. (2025), where $\lambda_R \approx 10^\circ\!\!.5$ was obtained in SFT simulations without surface inflows toward ARs. The sharp decline in the ratio with increasing $\lambda_R$ highlights a physical bifurcation in the solar dynamo's nonlinear response depending on the relative strength of transport processes. Models incorporating LQ show a broad range of contributions to the modulation of solar cycles, from insignificant to dominant (J. Jiang 2024). However, A. R. Yeates et al. (2025) found that, using historical data and SFT models, LQ has a stronger evidence base compared to TQ.

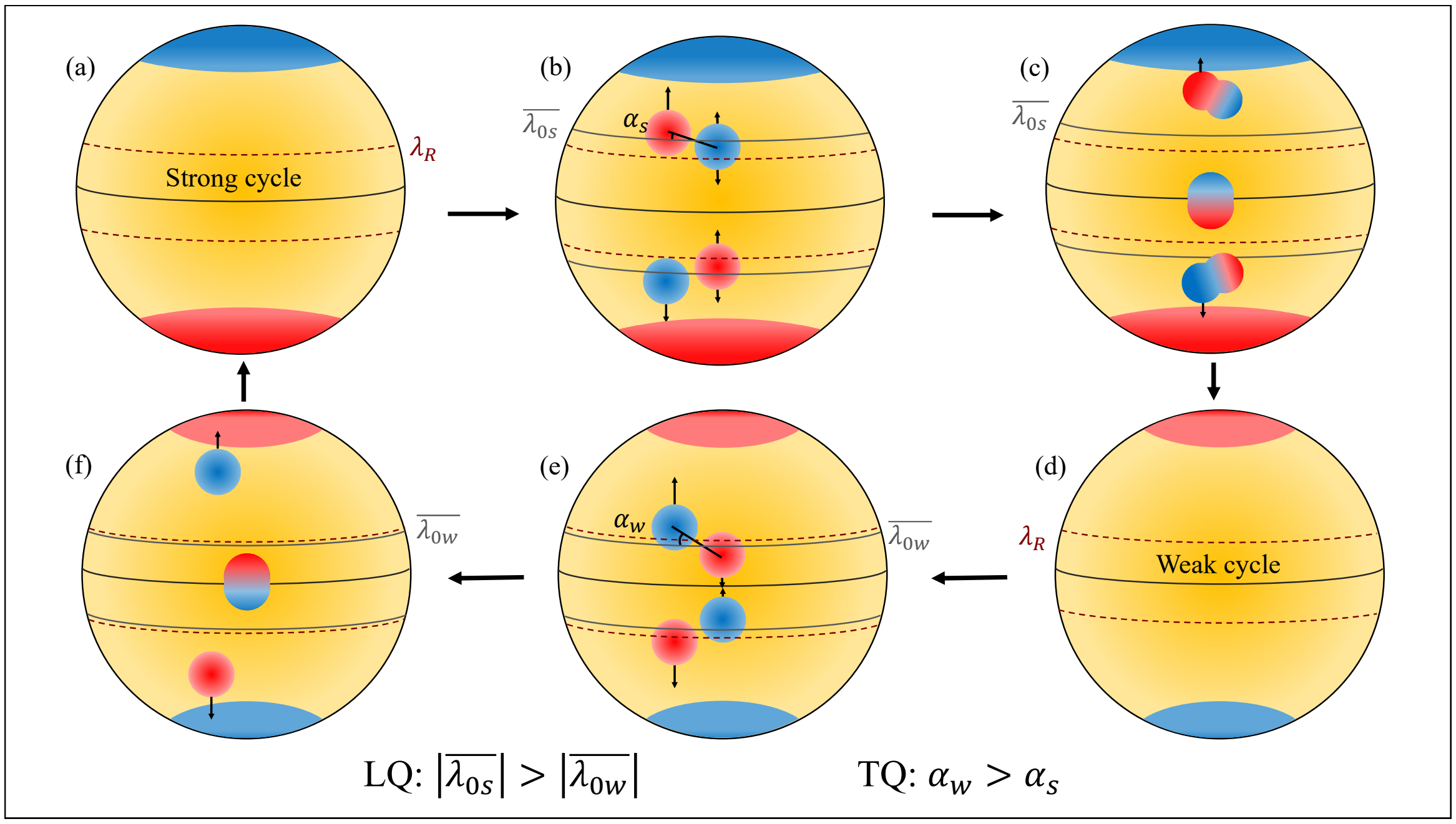

**Figure 4.** Schematic illustration of the roles of LQ and TQ in regulating solar cycle strength. Panels (a)–(c): strong cycle scenario. (a) Due to TQ, the mean tilt angle of emerging BMRs decreases from $\alpha_w$ (weak cycle) to $\alpha_s$ (strong cycle), enhancing flux cancellation between trailing and leading polarities. (b) LQ simultaneously shifts the mean emergence latitude poleward from $\bar{\lambda}_{0w}$ to $\bar{\lambda}_{0s}$, reducing cross-equatorial cancellation. (c) The combined effect decreases poleward transport of flux, leaving a weaker polar field and reducing the amplitude of the subsequent cycle (d). Panels (e), (f): weak cycle scenario. (e) With weaker TQ, BMRs retain larger Joy's law tilts, while LQ shifts the mean emergence latitude equatorward, enhancing cross-equatorial cancellation. (f) This facilitates stronger poleward transport of trailing flux, yielding an enhanced polar field that seeds a stronger following cycle (a). Together, these processes illustrate how the interplay of LQ and TQ can produce cycle-to-cycle modulation consistent with the Gnevyshev–Ohl rule and the observed even–odd alternation in solar activity.

We notice the nonlinearities have negligible dependence on the choice of $\tau$, as illustrated in Figure 3. This can be understood from the order-of magnitude scaling form of the SFT equation:

$$\frac{RB}{tu_0} \simeq \frac{B}{\lambda} + \left(\frac{\lambda_R}{\lambda}\right)^2 B - \frac{BR}{\tau u_0} + \frac{SR}{u_0}, \tag{18}$$

where the left-hand side is the characteristic temporal term and the terms on the right-hand side represent, respectively, advection, diffusion, radial decay, and source contributions (see Appendix B for derivation). Comparing the diffusion and decay terms, it becomes clear that the decay term is insignificant under realistic solar conditions. For typical surface diffusivity values ($\eta \approx 500\,{\rm km}^2\ {\rm s}^{-1}$), the decay timescale $\tau$ would need to be of the order of a few years or less to match the strength of the diffusive term. Such a small value is physically unrealistic, confirming that decay is not an efficient method for limiting the polar dipole moment when using PINN modulation. The PINN architecture enforces global smoothness during training, introducing an implicit decay-like regularization. This implicit effect naturally stabilizes the field without explicit exponential decay. This implies that the SFT solution is more dependent on $\lambda_R$ compared to $\tau$. Hence, nonlinearities such as TQ and LQ are more effective in regulating the polar dipole moment than the decay term.

Previous studies have explored the effect of nonlinearities on the solar dynamo; however, a clear physical interplay of the quenching mechanisms has not been discussed. We present a possible underlying physical operation of LQ and TQ in regulating the solar dipole moment for a given value of $\lambda_R$, schematically illustrated in Figure 4. During a strong cycle (see Figure 4(a)), due to TQ, the mean tilt angle of emerging bipolar magnetic regions (BMRs) reduces from its nominal value $\alpha_{\rm w}$ (in weak cycles) to a lower value $\alpha_{\rm s}$. This reduction enhances the possibility of flux cancellation between the trailing and leading opposite polarities. At the same time, LQ shifts the mean emergence latitude of BMRs from a lower value $\bar{\lambda}_{0\rm w}$ (in weak cycles) to a higher value $\bar{\lambda}_{0\rm s}$, pushing them farther away from the equator. $\lambda_R$ defines the equatorward limit of flux emergence that can be efficiently advected to the poles before diffusive decay dominates; consequently, BMRs emerging above this threshold experience longer diffusive timescales and weaker effective diffusion. As a result, cross-equatorial flux cancellation decreases once BMRs begin to emerge above $\lambda_R$ (see Figure 4(b)). The combined effect of changing tilt angle and mean emergence latitude due to TQ and LQ, respectively, results in a considerable increase in the flux cancellation between the leading and trailing polarities in a hemisphere. Consequently, the amount of magnetic flux transported toward the poles is decreased. This suppression of poleward flux transport significantly reduces the efficiency of poloidal field generation during strong cycles (see Figure 4(c)). As a result, stronger cycles contribute less effectively to the buildup of the dipole moment, ultimately leaving behind a weaker polar field and giving rise to a weaker subsequent cycle (Figure 4(d)).

In weak cycles, BMRs emerge with typical Joy's law tilts ($\alpha_w$), as TQ is weak, which lowers the probability of flux cancellation between the leading and trailing polarities. Simultaneously, LQ shifts the mean emergence latitude equatorward to $\bar{\lambda}_{0w}$, placing it below $\lambda_R$ where diffusive losses are more significant. Such a shift, further aided by surface inflows toward ARs (M. H. Talafha et al. 2025), enhances cross-equatorial flux cancellation (see Figure 4(e)). The combined effect results in less flux cancellation between the leading and trailing polarities within a hemisphere. As a result, the trailing polarity moves toward the poles, leading to a stronger polar field (see Figure 4(f)). The enhanced axial dipole at the end of a weak cycle then provides a stronger poloidal seed for the subsequent cycle, which in turn typically generates a stronger toroidal field and, consequently, a stronger solar cycle (Figure 4(a)). This is consistent with Gnevyshev–Ohl rule as discussed by P. Charbonneau et al. (2007) and Y. A. Nagovitsyn et al. (2024). This regulatory feedback is fundamental to the long-term stability of the solar cycle as it prevents the Sun from entering a state of unbounded cycle amplitude growth and promotes recovery from periods of very low activity.

The nonlinear feedback mechanism naturally implies an alternation between weak and strong solar cycles (i.e., an even–odd cycle pattern). However, such a perfect odd–even alternation in the actual solar cycle record has not been observed regularly. The classical expectation of an odd–even alternation rests on the simplifying assumption that the dynamo parameters remain constant over time. In reality, these parameters, particularly the dimensionless quantity $\lambda_R$, vary from cycle to cycle due to fluctuations in the meridional flow speed ($u_0$) and magnetic diffusivity ($\eta$). Consequently, the influence of LQ and TQ also changes between cycles. As a result, the system departs from the strict periodicity implied by the constant parameter scenario, producing the irregular sequence of strong and weak cycles as observed. This cycle-to-cycle modulation underscores the importance of incorporating time-dependent transport and diffusivity variations in dynamo models to capture the full complexity of solar cycle amplitude modulation.

The present study assumes a simplified time profile for the BMR source with constant model parameters and a strictly periodic cycle of 11 yr. While this facilitates clear comparison and the derivation of empirical trends, future work could benefit from incorporating cycle-to-cycle variations and in-cycle variation of SFT parameters to assess the impact of realistic variability on quenching efficiency and dipole moment buildup. Such a study could be extended to reconstruct the solar magnetic flux buildup as observed by various telescopes over the past several decades.

## 6. Conclusions

In this work, we examined the buildup of the solar dipole moment within a flux transport dynamo framework using PINN simulations. By introducing the residual dipole moment, defined as $D_t - e^{-P/\tau}D_{\text{cycle min}}$, we isolated the net amplification within each cycle. The sensitivity analysis of $\Delta D_{\text{LQ}}$ and $\Delta D_{\text{TQ}}$ clarified how the relative contributions from LQ and TQ vary across the dynamo parameter space. We found that $\Delta D_{\text{LQ}}$ decreases systematically with diffusivity $\eta$ but increases with meridional flow speed $u_0$, while $\Delta D_{\text{TQ}}$ remains comparatively insensitive to $\eta$ but grows modestly with $u_0$. The ratio $\Delta D_{\text{LQ}}/\Delta D_{\text{TQ}}$ highlights the transport-dependent balance between these two mechanisms. The weak sensitivity of the dipole evolution to $\tau$ in PINN modulation indicates that nonlinear feedback dominate over the decay term in constraining the buildup of the polar dipole moment.

We present a possible scenario of complex interplay between the LQ and TQ in regulating the solar dipole moment from cycle to cycle. The quenching mechanisms influence the poleward transport of magnetic flux, with the efficiency of this migration depending on the cycle strength and the underlying dynamic parameters (meridional flow and diffusion). This nonlinear interplay could naturally produce an alternation between weak and strong cycles, giving rise to the observed even–odd cycle pattern.

An important aspect of this study is the application of the PINN, which naturally embeds the governing equations into the learning process. Unlike conventional numerical schemes that rely on fine discretization and face stability constraints, PINNs achieve efficient and accurate representation of the solution across the full parameter space. This makes them especially well suited for capturing nonlinear feedback in the solar dynamo while remaining computationally efficient and adaptable for data-driven extensions.

Our results provide a framework for linking dynamo simulations with observations. Sunspot emergence latitudes, tilt angle catalogs, and synoptic magnetograms can serve as complementary probes of LQ and TQ, while polar field measurements provide direct validation of their impact. Future cycle-dependent analyzes of these datasets will allow the empirical testing and refinement of the quenching dependencies identified here, enabling data-constrained separation of LQ and TQ contributions to solar cycle variability.

In summary, nonlinear quenching mechanisms play a decisive role in regulating the solar dynamo and the buildup of the large-scale magnetic field. The interplay between LQ, TQ, and transport processes determines the effective range of dynamo operation and governs the strength of the polar field—the primary precursor of solar cycle predictability. By combining physical constraints with data-driven flexibility, the PINN offers a powerful pathway toward more realistic and robust dynamo modeling, with direct implications for forecasting solar cycle variability and space weather impacts. This framework paves the way for data-constrained PINN solar cycle forecasts, incorporating WSO synoptic magnetograms and upcoming high-resolution observations from DKIST, thereby bridging theoretical modeling with observational constraints.

## Acknowledgments

J.J.A. would like to express gratitude for the financial support received through the Prime Minister's Research

Fellowship. Part of this work was supported by the ESO under project No. Ts 17/2–1. Authors would also like to acknowledge the support of the ISRO RESPOND Grant (ISRO/RES/2/436/21-22).

*Software:* DeepXDE (L. Lu et al. 2021), NumPy (C. R. Harris et al. 2020), SciPy (P. Virtanen et al. 2020), and Matplotlib (J. D. Hunter 2007).

## Author Contributions

All authors contributed equally to this project.

## Appendix A
## Error Metrics

Let $y_i$ denote the observed (true) value, $\hat{y}_i$ the predicted value by the model, $\bar{y}$ the mean of the observed values, and $n$ the total number of data points.

$$\mathrm{MAE} = \frac{1}{n}\sum_{i=1}^{n}|y_i - \hat{y}_i|. \tag{A1}$$

$$\mathrm{MSE} = \frac{1}{n}\sum_{i=1}^{n}(y_i - \hat{y}_i)^2. \tag{A2}$$

$$\mathrm{RMSE} = \sqrt{\frac{1}{n}\sum_{i=1}^{n}(y_i - \hat{y}_i)^2}. \tag{A3}$$

$$R^2 = 1 - \frac{\sum_{i=1}^{n}(y_i - \hat{y}_i)^2}{\sum_{i=1}^{n}(y_i - \bar{y})^2}. \tag{A4}$$

$$\text{Max error} = \max_i |y_i - \hat{y}_i|. \tag{A5}$$

$$\mathrm{MAPE} = \frac{100\%}{n}\sum_{i=1}^{n}\left|\frac{y_i - \hat{y}_i}{y_i}\right|. \tag{A6}$$

The corresponding values of these error metrics are summarized in Table 2. The distribution of the error metrics for each model is shown in Figure 5.

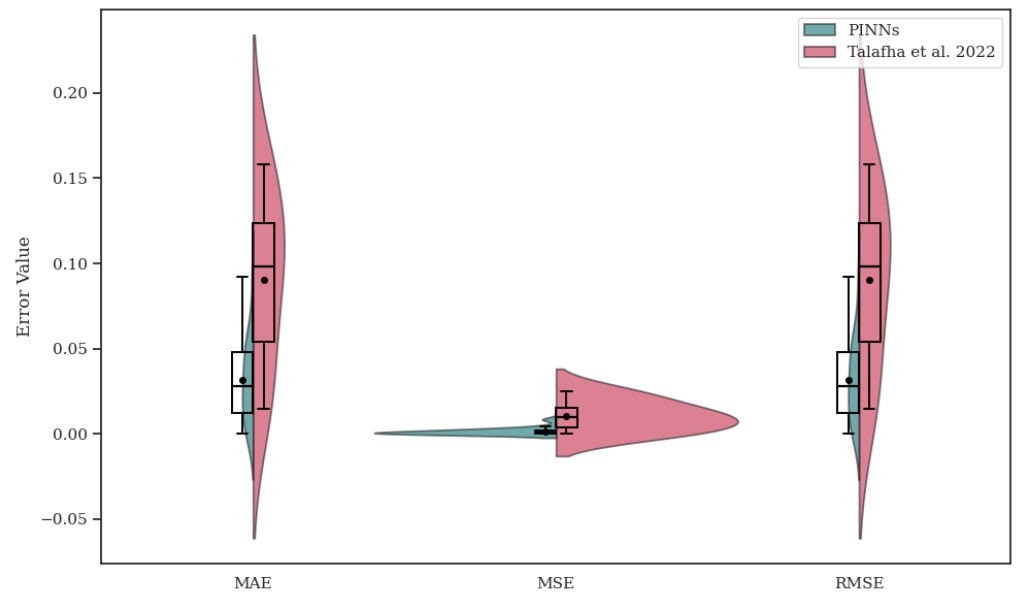

**Figure 5.** Distribution of errors for the PINNs model (teal) and Talafha et al. (crimson) across different metrics. Violin plots illustrate the full distributions of the errors, while overlaid box plots highlight the median (horizontal line), mean (black circle), and interquartile range (box) for each metric.

## Appendix B
## Order-of-magnitude Scaling of the SFT Equation

To obtain the scaling form of Equation (1), we perform a standard order-of-magnitude analysis using the characteristic quantities $B \sim B_*$, $t \sim t_*$, $u \sim u_0$, and $\partial/\partial\lambda \sim 1/\lambda$, where $\lambda$ denotes heliographic latitude in radians and $\cos\lambda = \mathcal{O}(1)$ over the activity belts. The corresponding characteristic scales of the various terms are

$$\frac{\partial B}{\partial t} \sim \frac{B_*}{t_*}, \quad \frac{1}{R}\frac{\partial(Bu)}{\partial\lambda} \sim \frac{u_0 B_*}{R\lambda},$$
$$\frac{\eta}{R^2}\frac{\partial^2 B}{\partial\lambda^2} \sim \frac{\eta B_*}{R^2\lambda^2}, \quad \frac{B}{\tau} \sim \frac{B_*}{\tau}.$$

Balancing these terms and multiplying by $R/u_0$ yields the nondimensional, order-of-magnitude form of the SFT equation,

$$\frac{R}{u_0}\frac{\partial B}{\partial t} \simeq \frac{B}{\lambda} + \left(\frac{\lambda_R}{\lambda}\right)^2 B - \frac{R}{u_0\,\tau}B + \frac{R}{u_0}S. \tag{B1}$$

Equation (B1) corresponds to the scaling relation used in Equation (18) of the main text. Here, the terms respectively quantify advection $(B/\lambda)$, diffusion $[(\lambda_R/\lambda)^2 B]$, decay $[BR/(u_0\tau)]$, and the source contribution $[SR/u_0]$. For typical solar parameters ($u_0 \sim$ 10–20 m s$^{-1}$, $\eta \sim$ 250–500 km$^2$ s$^{-1}$), one finds $\lambda_R \sim 7^\circ - 15^\circ$, indicating that the large-scale field evolution is primarily governed by the balance between advection and diffusion within the activity belts.

## ORCID iDs

Jithu J. Athalathil https://orcid.org/0009-0002-9114-5880
Mohammed H. Talafha https://orcid.org/0000-0003-4671-1759
Bhargav Vaidya https://orcid.org/0000-0001-5424-0059